\documentclass[12pt]{article}
\usepackage{amssymb}

\begin{document}

\title{Incompressible Quantum Hall Fluid}
\author{Bo-Yu Hou \thanks{Email:byhou@phy.nwu.edu.cn} and
        Dan-Tao Peng \thanks{Email:dtpeng@phy.nwu.edu.cn}\\
        Institute of Modern Physics, Northwest University\\
        Xi'an, 710069, P. R. China}
\date{}
\maketitle

\begin{abstract}
After review the quantum Hall effect on the fuzzy two-sphere
$S^2$ and Zhang and Hu's 4-sphere $S^4$, the incompressible
quantum Hall fluid on $S^2$, $S^4$ and torus are discussed
respectively. Next, the corresponding Laughlin wavefunctions on
$S^2$ are also given out. The ADHM construction on $S^4$ is
discussed. We also point out that on torus, the incompressible
quantum Hall fluid is related to the integrable Gaudin model and
the solution can be given out by the Yang Bethe ansatz.

\vspace{.5cm}


 \end{abstract}

\setcounter{equation}{0}

\section{Introduction}

\indent

Quantum Hall Effects (QHE) have been an important research
subject in condensed matter physics and theoretical physics.
A decade before,  Laughlin \cite{Laughlin} proposed the
incompressible quantum Hall fluid (QHF) formulation for
fractional quantization of the Hall effect (FQHF). Soon later
Haldane \cite{Haldane} considered a $2$ dimensional eletron gas
of $N$ particles moving on a spherical surface of radius $R$ in
a radial (Dirac monopole) magnetic field $B = \frac{\hbar S}{e
R^2}$, where $2 S$ is an integer as required by Dirac's
quantization condition. This describles a translational
(acturally rotational) invariant version of the incompressible
quantum Hall fluid (IQHF).

Last year, Susskind et al. constructed the noncommutative
Chern-Simons field theory for the QHF on ${\mathbb R}^2$
\cite{Suss} and $S^2$ \cite{BBST}. Zhang and Hu \cite{ZH, HZ}
generalized the QHE to a $4$ dimenstional sphere with
noncommutative geometry. In the ${\mathbb R}^2$ case,
Polychronokas \cite{Poly1} proposed a regularized version of the
noncommuative theory on the plane in the form of a finite
Chern-Simons matrix model with boundary field $\psi$
(equivalently a Wilson loop \cite{MP}). Hellerman and Raamsdonk
\cite{HR} suggested the coorespondent Laughlin-type
wavefunctions described by Chern-Simons matrix. Karabali and
Sakita \cite{KS1} gave an explicit form of Laughlin wavefunction
in terms of the energy eigenfunctions of Calogero model. These
works attract the interesting for QHE and QHF in connection with
the string and brane theory \cite{Fab, CHH}.

The purpose of this paper is to generalize the description of
the QHF which has been obtained on the noncommutative (N.C.)
plane ${\mathbb R}^2$ and fuzzy two-sphere to Zhang and Hu's
$4$-sphere and torus. In the next section, we first review the
Haldane's discription of the QHE on $2$-sphere, the
noncommutative algebra and the Moyal structure of the Hilbert
space is also shown in this section. After review Zhang and
Hu's generalization of the QHE to $4$-sphere, we give out the
noncommutative algera with the Moyal structure on fuzzy
$4$-sphere in section 3. In section 4, by using the method
suggested by Susskind \cite{Suss} and Polychronakos
\cite{Poly1, Poly2}, we construct the N.C. Chern-Simons theory
to descisbe the incompressible QHF on $S^2$ and give out the
algebra structure on $S^2$. The corresponding incompressible QHF
theory for Zhang and Hu's $4$-sphere is given in section 5. In
section 6, we construct the description of the incompressible
QHF on torus and relate it to the integrable Gaudin model which
can be solved by the Bethe-Yang ansatz. In the last section, we
shortly give some subjects which will be investigated later.

\section{The one particle wavefunctions in LLL on the first Hopf
fibration $S^2$}

\indent

On the two-dimensional sphere $S^2 = (S^3 \sim SU(2))/(S^1 \sim
U(1))$, the Hamiltonian of a single particle with charge $e$
moving around a Dirac monopele is
\begin{equation}
\label{Ham-S2}
H = \frac{|\vec{\Lambda}|^2}{2 M R^2} = \frac{\omega_C
|\vec{\Lambda}|^2}{2 \hbar S},
\end{equation}
here $M$ is the effective mass, $R$ is the raduis of $S^2$ and
$\omega_C = \frac{e B}{M}$ is the cyclotron frequency.
$\vec{\Lambda} = \vec{r} \times [ - i \hbar \bigtriangledown + e
\vec{a}(\vec{r})]$ is the orbital angular momentum with the
$U(1)$ gauge field $\vec{a}$ which the particle coupled with:
$\bigtriangledown \times \vec{a} = B \hat{r}$ and $B =
\frac{\hbar S}{e B^2}$ is the magnetic field strenth and
$\hat{r} = \frac{\vec{r}}{|r|}$. The commutation of the
dynamical angular momentum $\vec{\Lambda}$ is $[\Lambda^\alpha,
\Lambda^\beta] = i \hbar \epsilon^{\alpha \beta
\gamma}(\Lambda^\gamma - B \hat{r}^\gamma)$ which is not closed.
This Hamiltonian is rotation invariant and the corresponding
group manifold is $S^3$. The Euler angles $\alpha, \beta,
\gamma$ of $SU(2)$ is generated by the total angular momentum
$\vec{L}$:
\begin{equation}
\vec{L} = \vec{r} \times [- i \hbar \bigtriangledown + e
\vec{a}(\vec{r})] + \hbar S \frac{\vec{r}}{|r|} \equiv
\vec{\Lambda} + \hbar S \hat{r} = \vec{\Lambda} + \hbar \vec{S}
\end{equation}
which satisfies the commutation relations \cite{Haldane}
\begin{equation}
[L^\alpha, T^\beta] = i \hbar \epsilon^{\alpha \beta \gamma}
T^\gamma, \quad \mbox{$\vec{T} = \vec{L}, \vec{r}$ or
$\vec{\Lambda}$}
\end{equation}

The operators $\vec{S}^2$, $\vec{L}^2$, $L^3$ and
$\vec{\Lambda}^2$ can be diagonalized simultaneously and their
common eigenfuntions are
\begin{equation}
\Psi_{m, S}^J = D_{m, S}^J(\alpha, \beta, \gamma), \quad J \geq
|S|, m = -J, \cdots, J,
\end{equation}
here $D_{m, S}^J$'s are the finite rotation matrix elements and
the indices $J$, $m$ and $S$ are the eigenvalues of the
operators $\vec{L}^2$, $L^3$ and $\vec{S}$ respcetively. The
Eular angles $\omega = (\alpha, \beta, \gamma)$ equal $\alpha =
\phi, \beta = \theta$ and $\gamma = \gamma(\phi, \theta)$ which
has the $U(1)$ gauge freedom. In LLL ($J = S$ with energy
$\frac{1}{2} \hbar \omega_C$), these wavefunctions become
\begin{eqnarray}
\Psi_m^S & = & D_{m, -S}^S(\alpha, \beta, \gamma), \quad m = -S,
\cdots, S, \nonumber\\
& = & (- 1)^{S + m} \sqrt{\frac{2 S!}{(S + m)! (S - m)!}}u^{S -
m} v^{S + m}, \nonumber\\
& = & (1 + |\zeta|^2)^{- S} \zeta^{S - m} e^{i S \gamma}
\sqrt{\frac{2 S!}{(S + m)! (S - m)!}},
\end{eqnarray}
where the geodesic projection coordinate $\zeta =
\tan(\frac{\theta}{2}) \exp(- i \phi) = \frac{v}{u}$ and $u$,
$v$ are the spinor variables
\begin{eqnarray}
u & = & \cos \frac{\theta}{2} \exp(\frac{i \phi + i \gamma}{2})
= D_{\frac{1}{2}, - \frac{1}{2}}^{\frac{1}{2}}(\phi, \theta,
\gamma), \nonumber\\
v & = & \sin \frac{\theta}{2} \exp(\frac{- i \phi + i
\gamma}{2}} = D_{ \frac{1}{2}, \frac{1}{2}}^{\frac{1}{2}( \phi,
\theta, \gamma).
\end{eqnarray}

For this Hopf fibration (bundle) $S^2 = SU(2) / U(1)$, our base
space turns to be a K\"ahler manifold with the metric defined as
\begin{equation}
d s^2 = \frac{4 d \zeta d \bar{\zeta}}{(1 + |\zeta|^2)^2},
\end{equation}
and the corresponding symplectic structure (or the K\"ahler
form) is
\begin{equation}
\Omega = 2 i \frac{d \zeta \wedge d \bar{\zeta}}{(1 +
|\zeta|^2)^2} = 2 i \frac{\partial^2 K}{\partial \zeta \partial
\bar{\zeta}} d \zeta \wedge d \bar{\zeta},
\end{equation}
where $K = \ln (1 + |\zeta|^2)$ is the K\"ahler potential.

The Hilbert space ${\cal H}_N$ on this Hopf fibration $S^2$ is
composed by the $ N = 2 S + 1$ one particle wavefunctions
$\Psi_m^S$, $(m = -S, \cdots, S)$ around the Dirac monopole $g$
($S = g e$). The operators acting on these $2 S + 1$ states are
covariant $(2 S + 1) \times (2 S + 1)$ matrices, which as the
irreducible tensorial set should be (after normalized)
\begin{equation}
(X_M^J)_{m, m^\prime} = \left ( \begin{array}{ccc}
S & J & S \\
- m & M & m^\prime
\end{array} \right ), \quad 0 \leq J \leq 2 S, \quad 0 \leq M
\leq J,
\end{equation}
where $(:::)$ denotes the $3j$-symbol. These operators
constitute the right module \cite{HHY} of N. C. algebra ${\cal
A}_N$ on fuzzy sphere simply with the matrix product \cite{SW}:
\begin{equation}
\sum_m(X_{m_1}^{J_1})_{l m} (X_{m_2}^{J_2})_{m n} = \sum_{J, M}
\left ( \begin{array}{ccc}
J_1 & J_2 & J\\
m_1 & m_2 & M
\end{array} \right ) \left \{ \begin{array}{ccc}
J_1 & J_2 & J\\
S & S & S
\end{array} \right \} (X_M^J)_{l n},
\end{equation}
where $\{:::\}$ is the $6j$-symbol. This matrix product is
equivalent to the Moyal product on fuzzy sphere \cite{ARS}.
First, the coherent states which were found by Peremolov
\cite{Perelomov} on the coset space $J^2 = S^2$ in geodesic
projection coordinates are
\begin{equation}
| \omega ) \equiv | \theta, \phi ) = \sum_m D_{m, -S}^S(\phi,
\theta, -\phi) | S, m \rangle,
\end{equation}
which satisfy
\begin{equation}
\frac{N}{8 \pi^2} \int d \omega |\omega ) ( \omega | = 1, \quad
N = 2 S + 1.
\end{equation}
The Symbol of the operator $X_M^J$ is then defined as
\begin{eqnarray}
D_M^J(\theta, \phi) & \equiv & ( \omega | X_M^J | \omega ) =
D_{M, -J}^J(\phi, \theta, -\phi) \nonumber\\
& = & \sqrt{\frac{2 J!}{(J + M)! (J - M)!}}u^{J + M} v^{J - M}
\end{eqnarray}
and
\begin{equation}
X_M^J = \frac{2S + 1}{4 \pi} \int d \omega D_M^J(\theta, \phi) |
\omega(\theta, \phi) ) ( \omega(\theta, \phi) |.
\end{equation}
The symbol of the product of two operatos $X_{M_1}^{J_1}$ and
$X_{M_2}^{J_2}$ equals the "star product" (Moyar product) of two
symbols, namely
\begin{eqnarray}
D_{M_1}^{J_1} \star D_{M_2}^{J_2} & = &  \sum_{J, M} (-1)^{J_2 -
J_1 - M} (2 J + 1) \nonumber\\
& & \times \left ( \begin{array}{ccc}
J_1 & J_2 & J \\
M_1 & M_2 & M
\end{array} \right ) \left \{ \begin{array}{ccc}
J_1 & J_2 & J \\
S & S & S
\end{array} \right \} D_M^J.
\end{eqnarray}
This is a special case for the $\star$ product of the right
module of ${\cal A}_N$ on $S^2$ \cite{HHY}.

\section{LLL wave functions as the sections of the $2$nd Hopft
bundle}

\indent

On the $2$nd Hopf bundle (fibration), Zhang and Hu's $4$
dimensional fuzzy sphere $S^4 = (S^7 \sim Sp(4)/SU(2))/(S^3
\sim SU(2))$, which is equivalent to $SO(5) / (SO(3) \otimes
SO(3))$, the Hamiltonian of one particle is
\begin{equation}
\label{Ham-S4}
H = \frac{\hbar^2}{2 M R^2}\sum_{a < b} \Lambda_{a b}^2.
\end{equation}
This Hamiltonian is similar as (\ref{Ham-S2}) in the $S^2$ case,
but here the particle is coupling with a $SU(2)$ gauge field
$A_a$. $\Lambda_{a b}$ is the dynamical angular momentum given
by $\Lambda_{a b} = - i(x_a D_b - x_b D_a)$, where $D_a =
\partial_a + A_a$ is the covriant derivative and corresponding
field strength is $f_{a b} = [D_a, D_b]$. The symmetry of
$S^4$ is $SO(5)$, but $\Lambda_{a b}$ does not satisfy the
commutation rules of $SO(5)$. The one particle angular momentum
of the Yang's $SU(2)$ monopole is $L_{a b} = \Lambda_{a b} -
f_{a b}$ which obey the $SO(5)$ commutation rules.

The irreducible representaion of $SO(5)$ is labeled by two
integers $(r_1, r_2)$ with $r_1 \ge r_2 \ge 0$ and for such
representation, the Casimir operator and its
dimensionality are given by $C(r_1, r_2) = \sum_{a < b} L_{a
b}^2 = r_1^2 + r_2^2 + 3 r_1 + r_2$ and $D(r_1, r_2) =
\frac{1}{6}(1 + r_1 - r_2)(1 + 2 r_2)(2 + r_1 + r_2)(3 + 2 r_1)$
respectively. The $SU(2)$ gauge potential is valued by the
$SU(2)$ Lie algebra $[I_i, I_j] = i \epsilon_{i j k} I_k$ and
the corresponding Casimir is $\sum_i I_i^2 = I(I + 1)$ which
specifies the dimensions of the $SU(2)$ representation in the
monopole potential. So for a given $I$, from the Hamiltonian
(\ref{Ham-S4}), we can read out that the degeneracy of the
energy level is given by the dimensionality of the corresponding
irreducible representation $D(r_1, r_2)$.

The ground state, which is the lowest $SO(5)$ level labeled by
$(r_1 = I = \frac{p}{2}, r_2 = I = \frac{p}{2})$, is
$N = \frac{1}{6}(p + 1)(p + 2)(p + 3)$ fold degenrate, here $N$
also indicates the instanton number. The explicit expressions of
the  ground state wave functions of the Hilbert space ${\cal
H}_N$ in the spinor coordinates were given by Zhang and Hu in
Ref. \cite{ZH}:
\begin{equation}
\label{wavefun}
\Psi^{r_1, r_2}_{k_1, k_{1 z}; k_2, k_{2 z}; I, I_z}(\theta,
\alpha, \beta, \gamma, \alpha_I, \beta_I, \gamma_I) = \sqrt{
\frac{p !}{m_1 ! m_2 ! m_3 ! m_4 !}} \Psi_1^{m_1} \Psi_2^{m_2}
\Psi_3^{m_3} \Psi_4^{m_4},
\end{equation}
where $r_1 = r_2 = I = \frac{p}{2}$, $ p = m_1 + m_2 + m_3 +
m_4$ and $k_1 = \frac{m_3 + m_4}{2}, k_{1 z} = \frac{m_3 -
m_4}{2}, k_2 = \frac{m_1 + m_2 }{2}, k_{2 z} = \frac{m_1 -
m_2}{2}$; $k_i, k_{i z}$ $(i = 1, 2)$ are the eigenvalues of the
angular momentum of the stable (keeps $\vec{r}$ invariant)
subgroup $SO(3) \otimes SO(3)$ and $I, I_z$ are the eigenvalues
of the $SU(2)$ isopin.

The base space $S^4$ can be parameterized by the following
coordinate systems
\begin{eqnarray}
x_1 & = & \sin \theta \sin \frac{\beta}{2} \sin (\alpha -
\gamma), \nonumber\\
x_2 & = &  - \sin \theta \sin \frac{\beta}{2} \cos (\alpha -
\gamma), \nonumber\\
x_3 & = &  - \sin \theta \cos \frac{\beta}{2} \sin (\alpha +
\gamma), \\
x_4 & = & \sin \theta \cos \frac{\beta}{2} \cos (\alpha +
\gamma), \nonumber\\
x_5 & = & \cos \theta,
\end{eqnarray}
where $\theta, \beta \in [0, \pi)$ and $\alpha, \gamma \in [0, 2
\pi)$. Next, the orbital coordinates $x_a$ $(a = 1, \cdots, 5)$,
which describe the point on the $S^4$ by $X_a = R x_a$, is
related the spinor coordinates $\Psi_\alpha$ $(\alpha = 1, 2, 3,
4)$ through the relation
\begin{equation}
\label{x}
x_a = \bar{\Psi}\Gamma_a \Psi = \bar{\Psi}_\alpha
(\Gamma_a)_{\alpha \alpha^\prime} \Psi_{\alpha^\prime}, \quad
\sum_\alpha \bar{\Psi}_\alpha \Psi_\alpha = 1,
\end{equation}
where $\Gamma_a, (a = 1, \cdots, 5)$ are the five $4 \times 4$
Dirac matrices which satisfy the Cliford algebra $\{\Gamma_a,
\Gamma_b\} = 2 \delta_{a b}$. The explicit solution of
eq.(\ref{x}) is given by Zhang and Hu \cite{ZH}:
\begin{equation}
\left ( \begin{array}{c}
\Psi_1 \\
\Psi_2
\end{array} \right ) = \sqrt{\frac{1 + x_5}{2}} \left (
\begin{array}{c}
u_1 \\
u_2
\end{array} \right ), \quad \left ( \begin{array}{c}
\Psi_3 \\
\Psi_4
\end{array} \right ) = \sqrt{\frac{1}{2(1 + x_5)}}(x_4 - i x_i
\sigma_i) \left ( \begin{array}{c}
u_1 \\
u_2
\end{array} \right ),
\end{equation}
where $(u_1, u_2)$ is an arbitrary two components complex vector
satisfying $\sum_\sigma \bar{u}_\sigma u_\sigma = 1$ and it has
a rotation $SU(2)$ symmetry which maps to the same point $x_a$
on $S^4$. The corresponding group manifold is $S^3$ and the
direction of the $SU(2)$ isospin is specified by the Eular
angles $\alpha_I, \beta_I, \gamma_I$, which discribe the spinors
in $\Psi$:
\begin{equation}
u_1 = \cos \frac{\beta_I}{2} \exp(i \frac{\alpha_I +
\gamma_I}{2}), \quad u_2 = \sin \frac{\beta_I}{2} \exp(- i
\frac{\alpha_I - \gamma_I}{2}).
\end{equation}
The geometric connection gives out the $SU(2)$ gauge potential
$A_a$ of the Yang's monopole defined on $S^4$:
\begin{equation}
A_a = \frac{-i}{1 + x_5} \eta_{a b}^i x_b I_i, \quad \eta_{a
b}^i = \epsilon_{i a b 4} + \delta_{i a}\delta_{4 b} - \delta_{i
b}\delta_{4 a},
\end{equation}
where $I_i = \frac{\sigma_i}{2}$ and $\eta_{a b}^i$ is known as
the t'Hooft tensor.

By using the parameterization of $S^4$ and the choice of the
isospin coordinates, the one-particle wavefunction can be
rewritten as:
\begin{eqnarray}
\Psi^{(r_1, r_2)}_{k_1, k_{1 z}; k_2, k_{2 z}; r_2, \rho} & = &
(\sin \theta)^{-1} (1 - \cos \theta)^{k_1 + \frac{1}{2}} (1 +
\cos \theta)^{- k_2 - \frac{1}{2}} \nonumber\\
& & \times P^{k_1 + \frac{1}{2}, - k_2 - \frac{1}{2}}_{r_1 + 1 +
k_2 - k_1}(\cos \theta) U_{k_1, k_{1 z}; k_2, k_{2 z}; r_2,
\rho},
\end{eqnarray}
where $P^{\alpha, \beta}_n$ is the Jacobi polynomial and
\begin{equation}
U_{k_1, k_{1 z}; k_2, k_{2 z}; I_z, I_z^\prime} = \sum_{m, I_z}
\langle k, m; I, I_z | k_2, k_{2 z} \rangle D^{k_1}_{m, k_{1
z}}(\alpha, \beta, \gamma) D^I_{I_z, I_z^\prime}(\alpha_I,
\beta_I, \gamma_I) | I, I_z^\prime \rangle.
\end{equation}
These one-particle wavefunctions are the Yang's $SU(2)$ monopole
harmonics, i.e. the spherical harmonics on the coset space
$SO(5)/SU(2)$, which is locally isomorphic to the sphere
$S^4 \times S^3$.

The Hilbert space ${\cal H}_N$ on this second Hopf bundle
(fibration) $S^4$ is composed by all these one-particle
wavefunctons. The completeness of these one-paticle
wavefunctions is ensured by the following relation:
\begin{equation}
\sum_{p \ge r_1 \ge r_2 \ge 0} D[r_1, r_2] = (\frac{1}{6}(p +
1)(p + 2)(p + 3))^2 = N \times N,
\end{equation}
where $D[r_1, r_2]$ is the dimensionality the representation of
the $SU(2)$ gauge group given before which indicates the
degeracy of the energy level.

In the LLL level, using the one-particle wavefunctions as the
basis of the Hilbert space ${\cal H}_N$, the operators acting on
these $N$ states are covariant $N \times N$ matrices and they
can be also expressed by the Yang's $SU(2)$ monopole harmonics
$T^{(r_1, r_2)}_{k_1, k_{1 z}; k_2, k_{2 z}; r_2, \rho}$ and $p
\ge r_1 \ge r_2 \ge 0$, $J_1, J_2$ in Yang's lattice
$(\frac{p}{2}, \frac{q}{2}), (\frac{q}{2}, \frac{p}{2}),
(\frac{q}{2}, 0), (0, \frac{q}{2})$, $p = r_1 + r_2, q = r_1 -
r_2$, but herein the spinor $| I, I_z \rangle$ composed from
$u_1, u_2$ is replaced by operators $\Sigma^{r_2}_\rho$
fused from Pauli matrices $\sigma_i$. These operators also
consitute the right module of the N.C. algebra ${\cal A}_N$ on
$S^4$.

By making use of the integral formula about the Jacobi
polynomial and the product of the $SU(2)$ $D$-functions, the
matrix elements of the operator $T^R_J$ ($R = [r_1, r_2], J =
(j_1, j_{1 z}; j_2, j_{2 z}; r_2, \rho)$) are read as:
\begin{eqnarray}
\left \langle \begin{array}{c}
\frac{p}{2} \\
K^1
\end{array} \right | T^R_J \left |
\begin{array}{c}
\frac{p}{2} \\
K^2
\end{array} \right \rangle & = & \left \langle \begin{array}{c}
\frac{p}{2} \\
K^1
\end{array} \right | \left | \begin{array}{c}
R \\
J
\end{array} \right | \left | \begin{array}{c}
\frac{p}{2} \\
K^2
\end{array} \right \rangle \left \{ \begin{array}{ccc}
K_1^1 & j_1 & K_1^2 \\
I & r_2 & I \\
K_2^1 & j_2 & K_2^2
\end{array} \right \} \nonumber\\
& & \times \left ( \begin{array}{ccc}
K_1^1 & j_1 & K_1^2 \\
K_{1 z}^1 & j_{1 z} & K_{1 z}^2
\end{array} \right ) \left ( \begin{array}{ccc}
K_2^1 & j_2 & K_2^2 \\
K_{2 z}^1 & j_{2 z} & K_{2 z}^2
\end{array} \right ),
\end{eqnarray}
where $\left \langle \begin{array}{c}
\frac{p}{2} \\
K^1
\end{array} \right | \left | \begin{array}{c}
R \\
J
\end{array} \right | \left | \begin{array}{c}
\frac{p}{2} \\
K^2
\end{array} \right \rangle$ is independent of the $SU(2)$
magnetic quantum numbers, and it is expressed by the product of
two factors. One is the contribution of the integral with
respect to the variables $\alpha, \beta, \gamma, \alpha_I,
\beta_I$ and $\gamma_I$ and the other is given by the integral
part of $\theta$.

Similar as in the $S^2$ case, the Moyal product of the two
operatos $T^{R_1}_{J_1}$ and $T^{R_2}_{J_2}$ are also closed in
${\cal A}_N$:
\begin{equation}
T^{R_1}_{J_1} \star T^{R_2}_{J_1} = \sum_{R, J} \frac{1}{N}
\left ( \begin{array}{ccc}
R & R_1 & R_2 \\
J & J_1 & J_2
\end{array} \right ) \left \{ \begin{array}{ccc}
R & R_1 & R_2 \\
\frac{p}{2} & \frac{p}{2} & \frac{p}{2}
\end{array} \right \} T^R_J,
\end{equation}
here $R \equiv (r_1, r_2), J \equiv \left ( \begin{array}{cc}
J_1 & J_2 \\
J_{1 z} & J_{2 z}
\end{array} \right )$. This algebra is also a associate algbra.

\section{Two incompressible quantum Hall fluid on $S^2$}

\indent

When a D0-brane enters a D2-brane, it can dissolve into
magnetic flux, and its density is equivalent to a magnetic field
on the membrane while the D particle currents result in the
electric field. Through the T-duality, the D0-brane plays the
role of the electon which is the ends of strings on D2-brane
\cite{BBST, KaS, Berkooz, Brodie, BBO}. This give a
configuration of the branes and strings - the quantum Hall
soliton and the low energy dynamics display the fractional QHE
which can be modeled by a noncommutative Chern-Simons theory
\cite{BBST}.

In \cite{Suss}, Susskind propposed the non-commutative
Chern-Simons theory to describe the area preserving gauge
transformation of the incompressible electron gas on a constant
magnetic $B$ field. Here the area preserving gauge is:
\begin{equation}
X_i = x_i + \epsilon_{i j} \frac{\hat{A}_j}{2 \pi \rho_0},
\end{equation}
where $X_i$ is the noncommutative target space coordinate which
describes the positions of the electrons, $x_i$ is the comoving
coordinates of the incompressible electron gas. $\hat{A}_i$ is
the gauge field and $\theta = \frac{1}{2 \pi \rho_0}$ is the
noncommutative parameter. And the action of the noncommutative
Chern-Simons theory is:
\begin{equation}
S = \frac{k}{4 \pi} \int d^3 x \epsilon_{\mu \nu \rho}
\hat{A}_\mu \partial_\nu \hat{A}_\rho + \frac{2 i}{3}
\hat{A}_\mu \hat{A}_\nu \hat{A}_\rho,
\end{equation}
where $k \equiv B \theta = \frac{B}{2 \pi \rho_0} =
\frac{1}{\nu}$ and $\nu$ is called the filling fraction.

Up to a total divergent, this noncommutative Chern-Simons theory
is equivalent to the matrix model:
\begin{equation}
S = \int d t \frac{B}{2} {\rm Tr} \left \{ \epsilon_{a b}
(\dot{X}_a + i[A_0, X_a])X_b + 2 \theta A_0 \right \},
\end{equation}
here $X_a, (a = 1, 2)$ are two (infinite) target hermitian
"matrices" of the matrix theory \cite{Poly1}. [., .] is the
matrix commutator and ${\rm Tr}$ represents the (matrix) trace
over the Hilbert space. $A_0$ is the Lagrangian multiplier which
will generate the Gaussian constraint condition of the gauge
transformation. This action describes an incompressible fluid of
infinite particles on the plane in a constant magnetic field
$B$. The action of the finite $N$ particles system which
describes the dynamics of a Quantum Hall droplet is proposed by
Polychronakos \cite{Poly1}:
\begin{equation}
S = \int d t \frac{B}{2} {\rm Tr} \left \{ \epsilon_{a b}
(\dot{X}_a + i [A_0, X_a]) X_b + 2 \theta A_0 - \omega X_a^2
\right \} + \Psi^\dagger(i \dot{\Psi} - A_0 \Phi),
\end{equation}
here $X_a$ is represented by $N \times N$ matrices in matrix
theory. $\Psi$ is a complex $N$-vector which comes from the droplet
boundary states. The potential term $\omega X_a^2$ serves as a
spatial regulator and it breaks the translation invraiance of
the action.

In the fuzzy two-sphere $S^2$, the action is similar as the
droplet in ${\bf R}^2$:
\begin{equation}
\label{action}
S = \int d t \frac{B}{2} \left \{ \epsilon_{a b} (\dot{X}_a +
i [A_0, X_a]) X_b + 2 \theta A_0 \right \} + \Psi^\dagger (i
\Psi - A_0 \Psi).
\end{equation}
here $\Psi$ is a complex $N$-vector which defines the
$CP(N - 1)$ model manifold \cite{CCLY} and it has a left global
$U(N)$ symmetry and a right local $U(1)$ gauge symmetry which
can be mapped to the gauge field $\tilde{A}$ (statistical gauge
field) for D0 fluid by the Seiberg-Witten map. In this action,
the term $\frac{e B}{2} \epsilon_{a b} \dot{X}_a X_b$ is the
Lorentz force which gives out the Chern-Simons part in the
action as in plane ${\bf R}^2$ and $A_0$ is still the Lagrangian
multiplier.

The Gaussian constraint equation can be obtained by varying the
action (\ref{action}) with respect to the Lagrangian multiplier
$A_0$:
\begin{equation}
\label{G-constrain}
G \equiv = - i B [X_1, X_2] + \Psi \Psi^\dagger - B \theta = 0.
\end{equation}
The traceless part of this equation gives out the commutation
relation of $X_1$ and $X_2$ (moment map equation):
\begin{equation}
\label{momentmap}
[X_1, X_2] = i \theta (1 - |\Psi \rangle \langle \Psi |).
\end{equation}
This equation is the D-Flat equation in Supersymmetric
Yang-Mills theory. Here we have adopt an oversimplified
notation, i.e. in moving frame, for the $S^2$ without $\Psi$
source, the orbital augular momentum $\Lambda_i = \epsilon_{i
j} L_{j r}$ is equivalent to the Killing vector
$\bigtriangledown_j$ along $S^2$ surface:
\begin{eqnarray}
[\Lambda_+, \Lambda_-] & \simeq & J_r \equiv - S \sim B R^2 \sim
\theta. \nonumber\\
J_r |\Psi \rangle & = & S |\Psi \rangle, \quad |\Psi \rangle \in
{\cal H}_n.
\end{eqnarray}
For the right module ${\cal A}_n$ on $S^2$ without quasiparticle
\begin{eqnarray}
[X_\pm ,\star A] & \simeq & \bigtriangledown_\pm A, \quad A \in
{\cal A}_n, \nonumber\\
{[X_1, X_2]} & = & \theta, \quad X_\pm \equiv
\frac{1}{\sqrt{2}}(X_1 \pm i X_2),
\end{eqnarray}
here $\theta^2 \sim \frac{L^2}{r^2} \sim 2 S + 1 $ which is the
quantum effect.

In the plane case, the explicit expressions of the Laughlin
wavefunctions was given in \cite{HR}. Upon quantization, the
Gaussian constraint (\ref{G-constrain}) and the moment map
equation (\ref{momentmap}) require the physical states to be
singlets of $SU(n)$. The ground state being a completely
antisymmetric $SU(n)$ sigulet with $\Psi$ and $X_-$ has the
following form:
\begin{equation}
|\Phi_{\rm gr} \rangle = [\epsilon^{i_1 \cdots i_N}
\Psi_{i_1}^\dagger (\Psi^\dagger X_-)_{i_2} \cdots (\Psi^\dagger
X_{-}^{(N - 1)})_{i_N}]^k |0\rangle,
\end{equation}
where $X_{\pm} = \frac{1}{\sqrt{2}}(X_1 \pm i X_2)$ which
satisfy the constrain equation (\ref{G-constrain}) and
$|0\rangle$ is annihilated by $X_+$'s and $\Psi$'s, wihle the
excited states can be written as
\begin{equation}
|\Phi_{\rm exc} \rangle = \prod_{i = 1}^N ({\rm Tr}
X_-)^{c_i}[\epsilon^{i_1 \cdots i_N} \Psi^\dagger_{i_1}
(\Psi^\dagger
X_-)_{i_2} \cdots (\Psi^\dagger X_{-}^{(N - 1)})_{i_N} ]^k
|0\rangle,
\end{equation}

On the fuzzy two-sphere $S^2$, the Killing vectors on $S^2$
$\Lambda_\pm \sim X_\pm \sim D_{\mp 1}^1$ act on the state
$|\Phi\rangle \in H_n$ which is equivalent to the symbol $D_{M,
-S}^S$. The Hamiltonian of the system is
\begin{equation}
H = \frac{\omega_C (\hat{I}^2 - \hat{S}^2)}{2 \pi S}
\end{equation}

For the Incompressible Quantum Hall Fluid, the system now
contains $n$ particles and the moduli space is $S^{2 \otimes
n}/S_n$ with the hyperKa\"hler metric
\begin{equation}
d s^2 = \sum_{j = 1}^n \frac{d \zeta_j d \bar{\zeta}_j}{(1 +
|\zeta_j|^2)^2},
\end{equation}
and the simplectic form
\begin{equation}
\omega = \sum_{j = 1}^n 2 i \frac{d
\zeta_j \wedge d \bar{\zeta}_j}{(1 + |\zeta_j|^2)^2}.
\end{equation}

The ground state on $S^2$ is expressed by the Vandermonde
determinant of the spinor coordiantes $|\Psi_{i, j}| \simeq
\prod_{i < j} (u_i v_j - u_j v_i)$, where $\Psi_{i, j} \equiv
D_{i, -S}^S(\alpha_j, \beta_j, \gamma_j)$, while in the limit of
$S^2 \rightarrow {\mathbb C}_1$ plane, the quantum $(X_-)_{i j}
= \delta_{i j} \partial_i - \frac{i \theta}{\zeta_i - \zeta_j}(1
- \delta_{i j})$, $X_+ = {\rm diag}(z_1, \cdots, z_n)$, and the
excited state $|\Phi_{\rm exc}\rangle$ as the eigenfunction of
the {\bf Hamiltonian} $= {\rm Tr}(X_-^2)$ is expressed by Jack
polynomial.

\section{Incompressible quantum Hall fluid on $S^4$}

\indent

On the four-sphere $S^4$, we may set the brane construction as
$(2, 0)$ noncommutatity for electronic fluid or as IIA
supergravity for $D0$ fluid.

On the $S^4$, the four Killing vectors along the surface become
the basis $\Psi_i (i = 1, \cdots, 4)$, the fundamental $SU(2)$
monopole wave function s of C. N. Yang \cite{Yang1, Yang2},
which is conformally equivalent to the instanton on $4$-sphere.
So $S^4$ has a hyperK\"ahler symplectic metric. The left global
$n \times n$ quarternion matrix $(n = \frac{1}{6} s (s + 1)(s +
2)(s + 3))$ is the ADHM \cite{Atiyah} matrix for the normal
component of the projective module of instanton, with $n$ the
Pontrjagin number, i.e. instanton number. While the right local
$SU(2)$ gauge symmetry, in the bundle $SO(5)/SU(2) \sim$ base
$4$-sphere $SO(5)/(SU(2) \otimes SU(2)) \times$ gauge fibre
$SU(2)$ \cite{CHH} is the tangential part of this normal bundle,
i.e. determines the instanton potential. These are the same as
in Witten's paper \cite{W2}.

Now on $S^4$, the D-flat equation of noncommutative ADHM
constrains are:
\begin{eqnarray}
\mu_r & = & [B_1, B_1^\dagger] + [B_2, B_2^\dagger] + I
I^\dagger - J^\dagger J, \nonumber\\
\mu_c & = & I J + [B_1, B_2].
\end{eqnarray}

We introduce The 4 Killing vectors the
quarternion as
$$
q = \psi_\mu \sigma_\mu = \psi_1 {\bf 1} + \psi_2 {\rm\bf i} +
\psi_3 {\rm\bf j} + \psi_4 {\rm\bf k},
$$
here
\begin{eqnarray}
\psi_i & = & \psi_i({\bf r}_1, {\bf r}_2; {\bf k}_1, k_{1 z};
{\bf k_2}, k_{2 z}; {\bf s}, s_r) = D^{[\frac{1}{2},
\frac{1}{2}]}_{K_i}, \nonumber\\
\psi_1 & = & \psi(\frac{1}{2}, \frac{1}{2}; 0, 0; \frac{1}{2},
\frac{1}{2}; \frac{1}{2}, - \frac{1}{2}), \nonumber\\
\psi_2 & = & \psi(\frac{1}{2}, \frac{1}{2}; 0, 0; \frac{1}{2}, -
\frac{1}{2}; \frac{1}{2}, - \frac{1}{2}), \nonumber\\
\psi_3 & = & \psi(\frac{1}{2}, \frac{1}{2}; \frac{1}{2},
\frac{1}{2}; 0, 0; \frac{1}{2}, - \frac{1}{2}), \nonumber\\
\psi_4 & = & \psi(\frac{1}{2}, \frac{1}{2}; \frac{1}{2}, -
\frac{1}{2}; 0, 0; \frac{1}{2}, - \frac{1}{2}).
\end{eqnarray}
here $\psi({\bf r}_1, {\bf r}_2; {\bf k}_1, k_{1 z}; {\bf k}_2,
k_{2 z}; {\bf s}, s_r)$ is the eigenfunction of $SO(5)$
\cite{HZ, Yang1, Yang2}, $[{\bf r}_1, {\bf r}_2]$ are the $B_2$
($SO(5)$) Casimir; ${\bf k}_1$, ${\bf k}_2$ are the Casimir for
the two $SO(3)$ angular momentum in the stationary subgroup at
pole $SO(4) = SO(3) \otimes SO(3)$; $k_{1 z}$, $k_{2 z}$ are
their "magnetic" quantum number respectively; ${\bf s}$, $s_r$
are the (auti)selfdual $SU(2)$ gauge symmetry for $SU(2)$
monopole (instanton). $\psi_\mu$ is the Yang's $SU(2)$ monopole
harmonic function with $(p, q) = (1, 0)$, here $(p, q)$ is the
$Sp(2) \equiv C_2$ Casimir with $p = r_1 + r_2$, $q = r_1 -
r_2$. In the hedgehog gauge, it also represent the
(anti)selfdual stationary subgroup around the radial direction,
e.g. $s_r$ is that along $\hat{\bf r}$. In angular momentum
theory, the conventional rotation is $D_{{\bf k}, {\bf s}}^R$.
For the detail of $D$ function please confer to \cite{CHH}.

Now the Moduli space of the QHF on $S^4$ is $S^{4 \otimes
n}/S_n$. Next, we have the hyperK\"ahler symplectic form for
this noncommutative $S^4$ in terms of these Killing vectors:
\begin{equation}
\Omega = \sum_{i = 1}^n d q_i \wedge d \bar{q}_i,
\end{equation}
here the $q_i$ (the $\psi_{j i}$) depends on the shift of the
"center" of N.C. soliton from the north pole to $(\theta_i,
\alpha_i, \beta_i, \gamma_i)$.

The "area preserving" constrain equation with quasiparticle
source $I, J$ turns to be
\begin{equation}
[X, X^\dagger] = ({\rm\bf i} + {\rm\bf j} + {\rm\bf k})({\bf 1}
- n |v\rangle \langle v|),
\end{equation}
where $X = q + B = X_\mu \sigma_\mu, B = \left (
\begin{array}{cc}
- B_1 & B_2 \\
- B_2^\dagger & - B_1^\dagger
\end{array} \right )$, $({\bf i} + {\bf j} + {\bf k})$ are
self-dual orbital momentum and we have choosen $I, J$ to get an
Weyl invariant $\langle v| = \frac{1}{\sqrt{n}}(1, \cdots, 1)$.

\section{Incompressible Quantum Hall Fluid on Torus}.

\indent

On the torus ${\mathbb T}$, we choose the following comoving
frame coordinates of electrons:
\begin{equation}
z_i = \frac{1}{\sqrt{2}}(x_i + i y_i), \quad
\bar{z}_i = \frac{1}{\sqrt{2}}(x_i - i y_i), \quad (i = 1, 2,
\cdots, N).
\end{equation}
Then the moment map equation becomes
\begin{equation}
[z_j, \bar{z}_k]_\star = \frac{1}{B} \delta_{i j}, \quad
[\bar{z}_k,\star f(z)] = \partial_k f(z).
\end{equation}

As similar in the $S^2$ case, now on torus, the moduli space is
${\cal T}^{2 \otimes n}/S_n$ and corresponding symplectic form:
$\Omega = \sum_j d z_j \wedge d \bar{z}_j$. As a K\"ahler
manifold (with both complex and symplectic forms), the K\"ahler
potential is given by
\begin{equation}
K = \log \prod_{j \neq k} \sigma(z_j - z_k).
\end{equation}
Then the ground state wave function will be
\begin{equation}
\Psi_{\rm ground} = \prod_{j \neq k} \sigma(z_j - z_k).
\end{equation}

On the torus, there is an automorphism between the Wilson loop
algebra ${\cal A}_n$ and the Lie algebra $sl_n({\mathbb T})$
\cite{HPSY}. It is shown \cite{CFHSYY} that the level $l$
representation of the Lie algebra $sl_n({\mathbb T})$ on the
elliptic curve ${\mathbb T}$ is
\begin{eqnarray}
\label{E_alpha}
E_\alpha & = & (- 1)^{\alpha_1} \sigma_\alpha(0) \sum_j \prod_{k
\neq j} \frac{\sigma_\alpha(z_{j k})}{\sigma_0(z_{j k})} \left [
\frac{l}{n} \sum_{i \neq j} \frac{\sigma_\alpha(z_{j
i})}{\sigma_0(z_{j i})} - \partial_j \right ], \\
E_0 & = & - \sum_j \partial_j,
\end{eqnarray}
where $\alpha \equiv (\alpha_1, \alpha_2) \in Z_n \times Z_n$,
$\alpha \neq (0, 0)$, $z_{j k} = z_j - z_k$, $\partial_j =
\frac{\partial}{\partial z_j}$. $E_0$ commutes with $E_\alpha$.
In a more common basis, let $E_{i j} \equiv \sum_{\alpha \neq
(0, 0)} (I^\alpha)_{i j} E_\alpha$, where $(I_{\alpha_1,
\alpha_2})_{a b} = \delta_{a + \alpha_1, \alpha_2}
\omega^{b \alpha_2}$, we have
\begin{equation}
[E_{j k}, E_{l m}] = E_{j m} \delta_{k l} - E_{l k} \delta_{j
m},
\end{equation}
The Weyl reflection is realized by $j \leftrightarrow
k$ for $E_{j k} (j \neq k)$.

The D-flat equation with source then is
\begin{equation}
[D_z, \phi] = \zeta(1 - n | v \rangle \langle v |) \sigma^2(u),
\quad \langle v | = (1, \cdots, 1).
\end{equation}

The $sl(n)$ bundle, the Hamiltonian reduced by the moment
map with quasiparticle as a source, turns to be a
differential $L$ operator (quantum lax operator) of Gaudin model
\begin{equation}
L_{i j} = \sum_{\alpha \neq (0, 0)} w_\alpha(u) E_\alpha
(I_\alpha)_{i j}, \quad L_{i j} \sim \phi_{i j}
\end{equation}
where
\begin{equation}
w_\alpha(u) = \frac{\theta^\prime(0)
\sigma_\alpha(u)}{\sigma_\alpha(0) \sigma_0(u)}.
\end{equation}
The commutators of $L$ is
\begin{equation}
[L^{(1)}(u_1), L^{(2)}(u_2)] = [r^{(1, 2)}(u_1 - u_2),
L^{(1)}(u_1) \oplus L^{(2)}(u_2)].
\end{equation}
where the classical Yang-Baxter matrix
\begin{equation}
r(u)^{k, l}_{i, j} = \sum_{\alpha \neq (0, 0)} w_\alpha(u)
(I_\alpha)_i^k \otimes (I_\alpha^{-1})_j^l, \nonumber
\end{equation}
satisfies the classical Yang-Baxter equation
\begin{equation}
[r_{1 2}(u_{1 2}), r_{1 3}(u_{1 3})] + [r_{1 2}(u_{1 2}), r_{2
3}(u_{2 3})] + [r_{1 3}(u_{1 3}), r_{2 3}(u_{2 3}] = 0.
\end{equation}

The Hamitonian is then defined as
\begin{equation}
H = {\rm Tr}(L^2)
\end{equation}
and the Laughlin wave functions the eigenfunctions of this
Hamiltonian
\begin{equation}
{\rm Tr} (L^2) \psi = (- \partial^2_{z_i} + 2 \sum_i \wp(z -
z_i(t))) \psi = \epsilon \psi,
\end{equation}
where $\wp(z) = \partial^2 \sigma(z)$. This is a $q$ Lam\'e
equation and can be solved by the Bethe-Yang Ansatz.

\section{Discussion}

In this paper, we have succeeded in constructing the N.C.
algebra of the $S^2$ and $S^4$ and generalizing the
descriptions of the incompressible QHF to the fuzzy $4$-sphere
$S^4$ and torus ${\mathbb T}$. On the torus, we notice that the
incompressible QHF can be related to the integrable Gaudin model
which can be solved by the Bethe-Yang ansatz.

As seen in section 4, after T-duality, the D0 branes which enter
in D2 brane become the electrons which is the ends of the
strings on D2 brane. So for other noncommutative manifolds, it
is interesting to relate the matrix algebra given in this paper
with some applications in D-brane dynamics.

\end{document}